\documentclass{emulateapj}
\usepackage{apjfonts}
\usepackage{natbib,graphicx}

\def\wig#1{\mathrel{\hbox{\hbox to 0pt{%
          \lower.5ex\hbox{$\sim$}\hss}\raise.4ex\hbox{$#1$}}}}

\shorttitle{Discovery and Characterization of Kepler-12b}
\shortauthors{Fortney, et al.}

\newcommand{\mj}{$M_{\mathrm{J}}$}
\newcommand{\rj}{$R_{\mathrm{J}}$}
\newcommand{\me}{$M_{\oplus}$}

\newcommand{\rsun}{$R_{\odot}$}

\newcommand{\ms}{$M_{\odot}$}
\newcommand{\ko}{Kepler-12b}
\newcommand{\koi}{Kepler-12b}
\newcommand{\hd}{HD 209458b}

\newcommand{\te}{$T_{\rm eff}$}
\newcommand{\teff}{$T_{\rm eff}$}
\newcommand{\teq}{$T_{\rm eq}$}

\newcommand{\cp}{\citep}
\newcommand{\ct}{\citet}
\newcommand{\kk}{\emph{Kepler}}
\newcommand{\wspitzer}{\emph{Warm-Spitzer}}
\newcommand{\spitzer}{\emph{Spitzer}}
\newcommand{\kepler}{\emph{Kepler}}
\newcommand{\rhostar}{\ensuremath{\bar{\rho_\star}}}
\newcommand{\feh}{\ensuremath{\left[ {\rm Fe/H} \right]}} 

\slugcomment{Revised for ApJ, September 5, 2011}

\begin{document}

\title{Discovery and Atmospheric Characterization of Giant Planet Kepler-12b:\\An Inflated Radius Outlier}

\author{Jonathan J. Fortney\altaffilmark{1}$^{,}$\altaffilmark{21},
Brice-Olivier Demory\altaffilmark{2},
Jean-Michel D\'esert\altaffilmark{3},
Jason Rowe\altaffilmark{4},
Geoffrey W. Marcy\altaffilmark{5},
Howard Isaacson\altaffilmark{5},
Lars A. Buchhave\altaffilmark{6}$^{,}$\altaffilmark{3},
David Ciardi\altaffilmark{7},
Thomas N. Gautier\altaffilmark{8},
Natalie M. Batalha\altaffilmark{9},
Douglas A. Caldwell\altaffilmark{4},
Stephen T. Bryson\altaffilmark{10},
Philip Nutzman\altaffilmark{1},
Jon M. Jenkins\altaffilmark{4},
Andrew Howard\altaffilmark{5},
David Charbonneau\altaffilmark{3},
Heather A. Knutson\altaffilmark{5},
Steve B. Howell\altaffilmark{10},
Mark Everett\altaffilmark{11},
Fran\c{c}ois Fressin\altaffilmark{3},
Drake Deming\altaffilmark{12},
William J. Borucki\altaffilmark{10},
Timothy M. Brown\altaffilmark{13},
Eric B. Ford\altaffilmark{14},
Ronald L. Gilliland\altaffilmark{15},
David W. Latham\altaffilmark{3},
Neil Miller\altaffilmark{1},
Sara Seager\altaffilmark{2},
Debra A. Fischer\altaffilmark{16},
David Koch\altaffilmark{10},
Jack J. Lissauer\altaffilmark{10},
Michael R. Haas\altaffilmark{10},
Martin Still\altaffilmark{17},
Philip Lucas\altaffilmark{18},
Michael Gillon\altaffilmark{19}$^{,}$\altaffilmark{20},
Jessie L. Christiansen\altaffilmark{4},
John C. Geary\altaffilmark{3}
}

\altaffiltext{1}{Department of Astronomy and Astrophysics, University of California, Santa Cruz, CA 95064; jfortney@ucolick.org}
\altaffiltext{2}{Massachusetts Institute of Technology, Cambridge, MA 02139, USA}
\altaffiltext{3}{Harvard-Smithsonian Center for Astrophysics, 60 Garden Street, Cambridge, MA 02138}
\altaffiltext{4}{SETI Institute/NASA Ames Research Center, Moffett Field, CA 94035}
\altaffiltext{5}{Department of Astronomy, University of California, Berkeley, CA 94720-3411, USA} 
\altaffiltext{6}{Niels Bohr Institute \& StarPlan, University of Copenhagen, Denmark}
\altaffiltext{7}{NASA Exoplanet Science Institute, Caltech, MS 100-22, 770 South Wilson Avenue, Pasadena, CA 91125, USA}
\altaffiltext{8}{Jet Propulsion Laboratory/California Institute of Technology, Pasadena, CA 91109, USA}
\altaffiltext{9}{San Jose State University, San Jose, CA 95192, USA}
\altaffiltext{10}{NASA Ames Research Center, Moffett Field, CA 94035}
\altaffiltext{11}{National Optical Astronomy Observatories, Tucson, AZ 85719}
\altaffiltext{12}{Department of Astronomy, University of Maryland at College Park, College Park, MD 20742}
\altaffiltext{13}{Las Cumbres Observatory Global Telescope, Goleta, CA 93117}
\altaffiltext{14}{University of Florida, Gainesville, FL 32611}
\altaffiltext{15}{Space Telescope Science Institute, Baltimore, MD 21218}
\altaffiltext{16}{Yale University, New Haven, CT 06520 US}
\altaffiltext{17}{Bay Area Environmental Research Institute/NASA Ames Research Center, Moffett Field, CA 94035, USA}
\altaffiltext{18}{Centre for Astrophysics Research, University of Hertfordshire, College Lane, Hatfield, AL10 9AB, England}
\altaffiltext{19}{Institut d'Astrophysique et de Géophysique, Université de Liège, Allée du 6 Août 17, Bat. B5C, 4000 Liège, Belgium}
\altaffiltext{20}{Observatoire de Genève, Université de Genève, 51 Chemin des Maillettes, 1290 Sauverny, Switzerland}
\altaffiltext{21}{Alfred P. Sloan Research Fellow}

\begin{abstract}
We report the discovery of planet \ko\ (KOI-20), which at $1.695\pm0.030$ \rj\ is among the handful of planets with super-inflated radii above 1.65 \rj.  Orbiting its slightly evolved G0 host with a 4.438-day period, this $0.431 \pm 0.041$ \mj\ planet is the least-irradiated within this largest-planet-radius group, which has important implications for planetary physics.  The planet's inflated radius and low mass lead to a very low density of $0.111 \pm 0.010$ \mbox{g cm$^{-3}$}.  We detect the occultation of the planet at a significance of 3.7$\sigma$ in the \emph{Kepler} bandpass.  This yields a geometric albedo of $0.14\pm0.04$; the planetary flux is due to a combination of scattered light and emitted thermal flux.  We use multiple observations with Warm \emph{Spitzer} to detect the occultation at 7$\sigma$ and 4$\sigma$ in the 3.6 and 4.5 $\mu$m bandpasses, respectively.  The occultation photometry timing is consistent with a circular orbit, at $e<0.01$ (1$\sigma$), and $e<0.09$ (3$\sigma$).  The occultation detections across the three bands favor an atmospheric model with no dayside temperature inversion.  The \kk\ occultation detection provides significant leverage, but conclusions regarding temperature structure are preliminary, given our ignorance of opacity sources at optical wavelengths in hot Jupiter atmospheres.  If \ko\ and \hd, which intercept similar incident stellar fluxes, have the same heavy element masses, the interior energy source needed to explain the large radius of \ko\ is three times larger than that of \hd.  This may suggest that more than one radius-inflation mechanism is at work for \ko, or that it is less heavy-element rich than other transiting planets.
\end{abstract}


\section{Introduction}
Transiting planets represent an opportunity to understand the physics of diverse classes of planets, including mass-radius regimes not found in the solar system.  The knowledge of the mass and radius of an object immediately yields the bulk density, which can be compared to models to yield insight into the planet's internal composition, temperature, and structure \cp[e.g.,][]{Miller11}.  Subsequent observations, at the time of the planet's occultation (secondary eclipse) allow for the detection of light emitted or scattered by the planet's atmosphere, which can give clues to a planet's dayside temperature structure and chemistry \cp{Marley07b,Seager10}.  NASA's \kk\ Mission was launched on 7 March 2009 with the goal of finding Earth-sized planets in Earth-like orbits around Sun-like stars \cp{Borucki10}.  While working towards this multi-year goal, it is also finding an interesting menagerie of larger and hotter planets that are aiding our understanding of planetary physics.

Early on in the mission, followup radial velocity resources preferentially went to giant planets, for which it would be relatively easy to confirm their planetary nature through a measurement of planetary mass.  This is how the confirmation of planet \ko\ was made, at first glance a relatively standard ``hot Jupiter'' in a 4.438 day orbit.  However, upon further inspection, the mass and radius of \ko\ make it an interesting planet from the standpoint of the now-familiar ``radius anomaly'' of transiting giant planets \cp[e.g.][]{Charb06,Burrows07,Laughlin11}.  Given our current understanding of strongly-irradiated giant planet thermal evolution, around 1/3 to 1/2 of known transiting planets are larger than models predict for several-Gyr-old planets that cool and contract under intense stellar irradiation \cp{Miller09}.

The observation that many Jupiter- and Saturn-mass planets are be larger than 1.0 Jupiter-radii can be readily understood. It is the magnitude of the effect that still needs explanation.  The first models of strongly irradiated planets yielded the prediction that these close-in planets would be inflated in radius compared to Jupiter and Saturn \cp{Guillot96}.  The high incident flux drives the radiative convective boundary from less than a bar, as in Jupiter, to pressures near a kilobar.  The thick radiative zone transports less flux than a fully convective atmosphere, thereby slowing interior cooling, which slows contraction.  A fairly uniform prediction of these strongly irradiated models is that $1.2-1.3$ \rj\ is about the largest radii predicted for planets several gigayears-old \cp{Bodenheimer03,Burrows07,Fortney07a,Baraffe08}.  However, planets commonly exceed this value.

\begin{figure}
\plotone{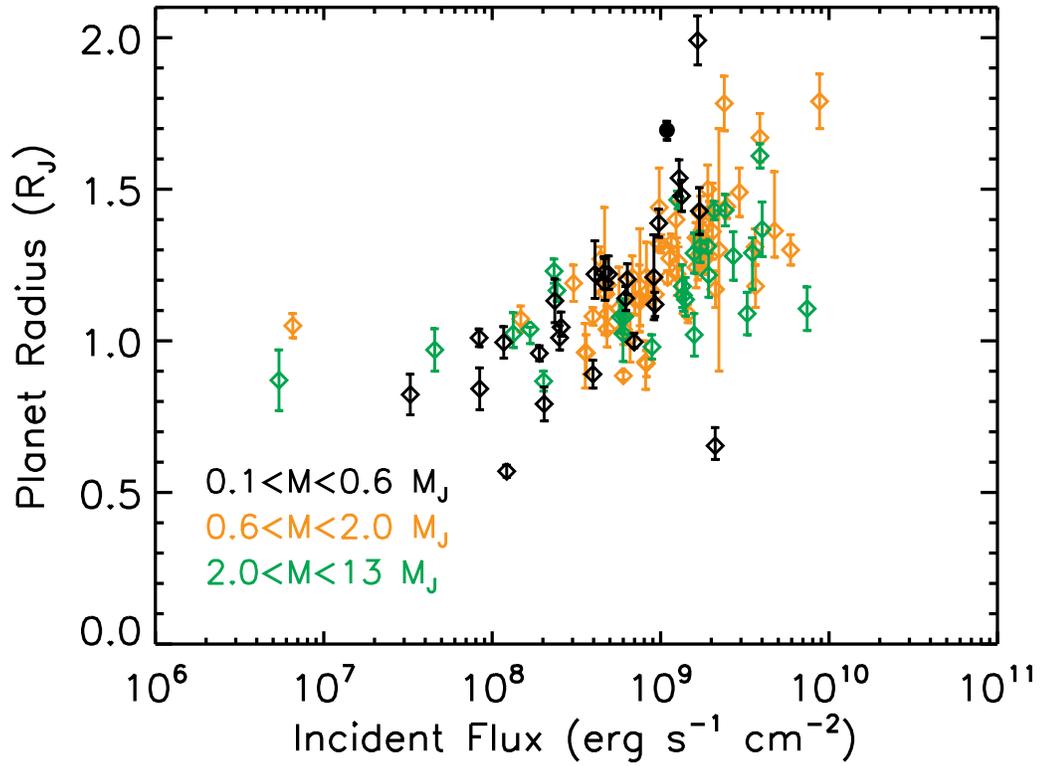}
\caption{Observed planetary radius as a function of total incident flux for the known transiting gas giants.  Planets are plotted in three colors for three different mass cuts.  \ko\ is shown as a black filled circle.  Note the general trend towards smaller radii with decreased insolation.   There is a probable break in slope at incident fluxes of $1-2\sim 10^8$erg cm$^{-2}$ s$^{-1}$.    Planets are taken from the compilation at: http://www.inscience.ch/transits/.
\label{rvsi}}
\end{figure}

The mechanism that leads to the radius anomaly has not yet been definitively identified.  However, constraints are emerging.  One is planet radius vs.~incident flux, which could also be thought of as radius vs.~equilibrium temperature, with an assumption regarding planetary Bond albedos.   Figure \ref{rvsi} shows planet radii vs.~incident flux for the transiting systems with confirmed masses.  Since low-mass planets are relatively easier to inflate to large radii than higher mass planets \cp[e.g.][]{Miller09}, we plot the planets in three mass bins.  The lowest mass bin is Saturn-like masses, while the middle mass bin is Jupiter-like masses.  The upper mass bin ends at 13 \mj, the deuterium burning limit.  Kepler-12b is shown as a black filled circle.  The largest radius planets are generally the most highly irradiated \cp{Kovacs10,Laughlin11,Batygin11}.  The near-universality of the inflation, especially at high incident fluxes, now clearly argues for a mechanism that affects all close-in planets \cp{Fortney06}, rather than one that affects only some planets.  The distribution of the radii could then be understood in terms of differing magnitudes of the inflation mechanism, together with different abundances of heavy elements within the planets \cp{Fortney06,Guillot06,Burrows07,Miller11,Batygin11}.

Within this emerging picture, outlier points are particularly interesting:  those that are especially large, given their incident flux.  These are the super-inflated planets with radii of 1.7 \rj\ or larger.  These include WASP-12b \cp{Hebb09}, TrES-4b \cp{Mandushev07,Sozzetti09}, WASP-17b \cp{Anderson10}, and now \ko, which is the least irradiated of the four.  In the following we describe the discovery of \ko, along with the initial characterization of the planet's atmosphere.

Transiting planets enable the characterization of exoplanet atmospheres.  The \emph{Spitzer Space Telescope} has been especially useful for probing the dayside temperature structure of close-in planetary atmospheres, as thermal emission from the planets can readily be detected by \emph{Spitzer} at wavelengths longer than 3 $\mu$m.  Data sets are becoming large enough that one can begin to search for correlations in the current detections \cp{Knutson10,Cowan11}.

A powerful new constraint of the past two years is the possibility of joint constraints in the infrared, from \emph{Spitzer}, and the optical, from space telescopes like \emph{CoRoT} \cp[e.g.,][]{Gillon10,deming11} and \kk\ \cp{desert11b}.  The leverage from optical wavelengths comes from a measurement (or upper limit) of the geometric albedo of the planet's atmosphere, although this is complicated by a mix of thermal emission and scattered light both contributing for these planets.  Detection of relatively low geometric albedos $A_{\mathrm g}<0.15$ is consistent with cloud-free models of hot Jupiter atmospheres \cp{Sudar03,Burrows08b}, and can inform our understanding of what causes the temperature inversions in many hot Jupiter atmospheres \cp{Spiegel10b}.

In this paper we discuss all aspects of the detection, validation, confirmation, and characterization of the planet.  Section \ref{discov} discusses the detection of the planet by \kk, while \S \ref{confirm} covers false-positive rejection and radial velocity confirmation.  Section \ref{secparams} gives the global fit to all data sets to derive stellar and planetary parameters, while \S \ref{2nd} concerns the observational and modeling aspects of atmospheric characterization.  Section \ref{discuss} is a discussion of the planet's inflated radius amongst its peers, while \S \ref{conclu} gives our conclusions.

\section{Discovery} \label{discov}

The {\it Kepler} science data for the primary transit search mission
are the long cadence data \cp{Jenkins10a}. These consist of
sums close to 30 minutes of each pixel in the aperture containing the
target star in question. These data proceed through an analysis
pipeline to produce corrected pixel data, then simple unweighted
aperture photometry sums are formed to produce a photometric
time series for each object \cp{Jenkins10b}. The many
thousands of photometric time series are then processed by the
transiting planet search (TPS) pipeline element \cp{Jenkins10b}.

\begin{figure}
\plotone{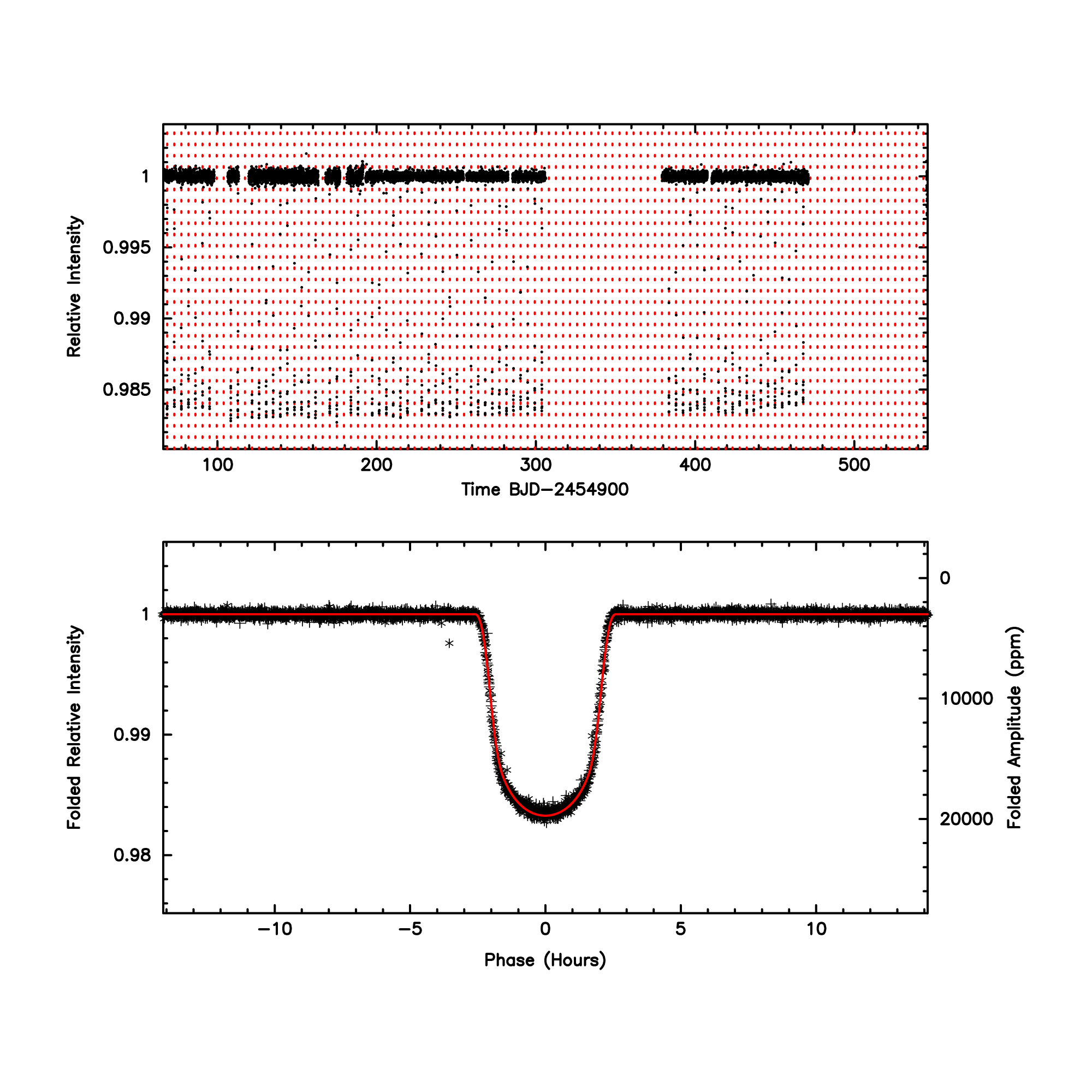}
\caption{Time series and folded transit light curve for \koi.
\label{lc}}
\end{figure}

The candidate transit events identified by TPS are also
vetted by visual inspection.
The light curves produced by the photometry pipeline tend
to show drifts due to an extremely small, slow focus change \cp{Jenkins10b}, and there are also sometimes low frequency
variations in the stellar signal that can make analysis
of the transit somewhat problematic. These low-frequency
effects can be removed by modest filters that have only an
insignificant effect on the transit signal \cp{Koch10b}.
The unfolded and folded light curves for \ko\ produced in
this manner are shown in Figure \ref{lc}.

Centroid analysis was performed using both difference image \cp{Torres11} and photocenter motion \cp{Jenkins10c} techniques using Q1 through Q4 data.  This analysis indicates that the object with the transiting signal is within 0.01 pixels (0.04 arcsec) of Kepler-12, which is the $3 \sigma$ radius of confusion (including systematic biases) for these techniques.  

The parent star, {\it Kepler} Input Catalog (KIC) identification number 11804465, has a magnitude in the \kk\ band of 13.438.   The KIC used ground-based multi-band photometry to assign an effective temperature and surface gravity of \te\ = 6012 K and log $g$ = 4.47 (cgs) to Kepler-12, corresponding to a late-F or early-G dwarf. Stellar gravities in this part of the H-R diagram are difficult to determine from photometry alone, and one of our conclusions based on high-resolution spectroscopy and light curve analyses in \S 4 is that the star is near
the end of its main-sequence lifetime, with a radius that has expanded to $R_* = 1.483 \pm 0.027$ \rsun\ and a surface gravity of log $g = 4.175 \pm 0.013$. In turn, this implies
an inflated radius for the planet candidate, originally known as Kepler Object of Interest (KOI)-20 \cp{Borucki11}. This conclusion is hard to avoid,
because the relatively long duration of the transit, more than 5 hr
from first to last contact, demands a low density and expanded
radius for the star.

\section{Confirmation: Follow-up Observations} \label{confirm}
\subsection{High Resolution Imaging from Large Telescopes}
\begin{figure} \epsscale{0.75}
\plotone{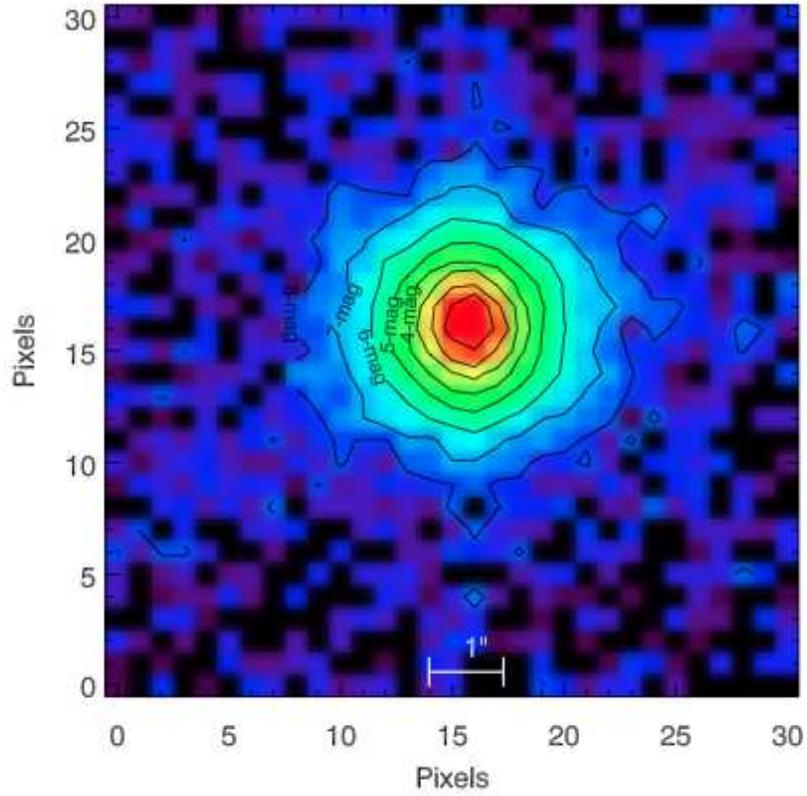}
\caption{Image of Kepler-12 taken with the Keck 1 telescope guide camera, showing $9\times9$ arcsec taken in 0.8 arcsec seeing.   North is up and east is to the left and the pixels are 0.30 arcsec in size.  The exposure time was 1.0 sec.  The detector is a Photometrics CCD and the filter is a BG38, making the passband roughly 400 - 800 nm, similar to that of {\it Kepler}.   Contours show surface brightness relative to the core.   No companion is seen down to 7 magnitudes fainter than Kepler-12 beyond $\sim$1 arcsec from it.  There is no evidence of a star that could be an eclipsing binary.
\label{keck}}
\end{figure}

Blends due to unresolved stellar companions (associated or background) can only be ruled out with direct imaging from large telescopes.  In Figure \ref{keck} we show an image of Kepler-12 taken with the Keck I telescope guide camera, showing $9\times9$ arcsec taken in 0.8 arcsec seeing.  This 1.0 second exposure was taken with a  BG38 filter, making the passband roughly 400 - 800 nm, similar to that of {\it Kepler}.   Contours show surface brightness relative to the core.   No companion is seen down to 7 magnitudes fainter than Kepler-12 beyond $\sim$1 arcsec from it.  Thus, there is no evidence of a star that could be an eclipsing binary, consistent with the lack of astrometric displacement during transit.

In addition, speckle observations using the WIYN telescope were made on the night of 18/19 June 2010, as part of the \emph{Kepler} followup program of S.~Howell and collaborators \cp{howell11}.  No additional source were seen to 3.69 magnitudes fainter in $R$-band and 2.17 magnitudes fainter in $V$-band in an annulus around the star spanning between 0.1-0.3 arcsec in radius.  No companions could be seen as close as the diffraction limit (0.05 arcsec from the star) or as far as the edge of the $2.8\times2.8$ arcsecond FOV.

\begin{figure} \epsscale{0.75} 
\includegraphics[angle=90,width=6.5in]{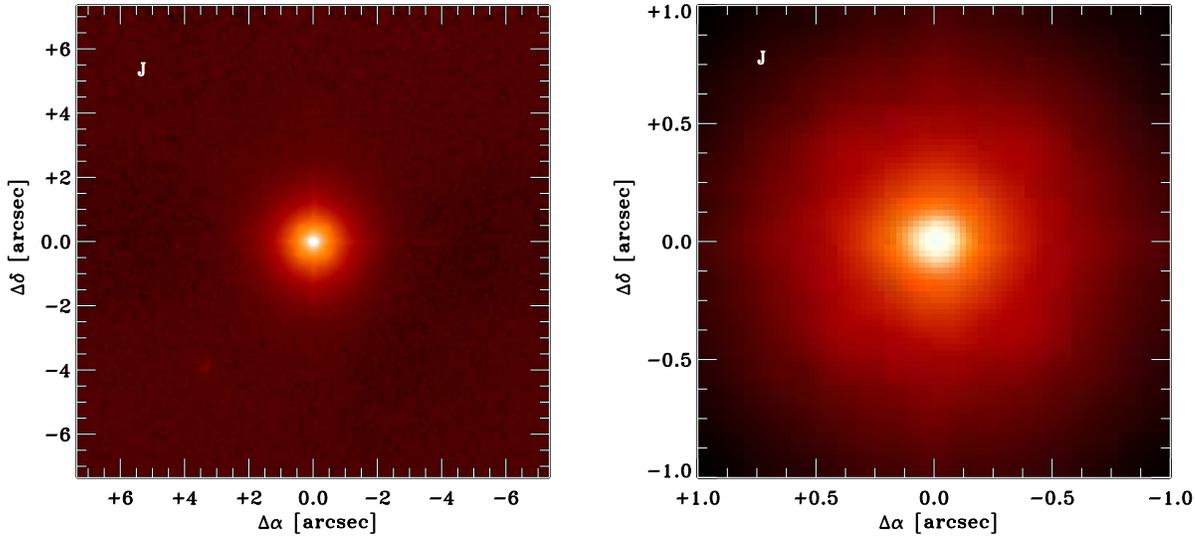}
\caption{$J$ Palomar adaptive optics image of Kepler-12.  The left image displays a
$15^{\prime\prime} \times 15^{\prime\prime}$ field of view centered on the
primary target.  The right image displays a  $2^{\prime\prime} \times
2^{\prime\prime}$ field of view centered on the primary target. The four-point
pattern surrounding the central point spread function core is part of the
adaptive optics point spread function.
\label{palo}}
\end{figure}

Near-infrared adaptive optics imaging of Kepler-12 was obtained on the
night of 08 September 2009 UT with the Palomar Hale 200in telescope and the
PHARO near-infrared camera \citep{hayward2001} behind the Palomar adaptive
optics system \citep{troy2000}.  PHARO, a $1024\times20124$ HgCdTe infrared
array, was utilized in the 25.1 mas/pixel mode yielding a field of view of
25$^{\prime\prime}$.  Observations were performed in $J$ filter ($\lambda_0
= 1.25\mu$m).  The data were collected in a standard 5-point quincunx
dither pattern of 5$^{\prime\prime}$ steps interlaced with an off-source
(60$^{\prime\prime}$ East) sky dither pattern. Data were taken with
integration times per frame of 60 sec (15 frames) for a total on-source
integration time of 15 minutes.  The individual frames were reduced with a
custom set of IDL routines written for the PHARO camera and were combined
into a single final image.  The adaptive optics system guided on the
primary target itself and produced  a central  core width of $FWHM =
0.11^{\prime\prime}$.  The final coadded image at $J$ is shown in
Figure~\ref{palo}.

\begin{figure} \epsscale{0.75} 
\plotone{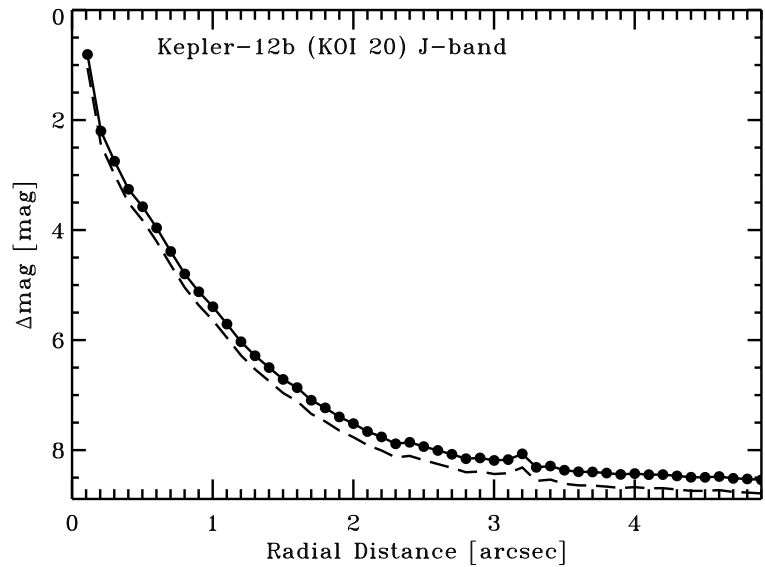}
\caption{Palomar detection limits as a function of radial distance from the primary target, Kepler-12.  The filled circles represent the J-band limits and each point represents a step in FWHM away from the primary target centroid peak.  The dashed line underneath represents the J-band limits converted to \emph{Kepler} magnitude limits if a star were to have a nominal $m_{Kepler}-J$ color. (For a magnitude limited sample, the median $m_{Kepler}-J$ = $1.28\pm0.52$).
\label{palo-sen}}
\end{figure}

One additional source was detected at 5$^{\prime\prime}$ SE and $\Delta J
\approx 8$ magnitudes fainter than the primary target, near the limit of
the observations.   No additional sources were detected at $J$ within $7
\farcs 5$ of the primary target.  Source detection completeness was
evaluated by measuring the median level and dispersion within a series
of annular rings, surrounding the primary target.  Each ring has a width of
$0.11^{\prime\prime} = 1$ FWHM, and each successive ring is stepped from
the previous ring by $0.11^{\prime\prime} = 1$ FWHM.  The median flux level
and the dispersion of the individual rings were used to set the $4\sigma$
sensitivity limit within each ring.  The measured limits are in the
$J$-band, but have been converted to limits in the \emph{Kepler} bandpass based
upon the typical $m_{Kepler}-J = 1.28 \pm 0.52$ mag for a magnitude limited
sample \citep{howell11}.  A summary of the detection efficiency as a
function of distance from the primary star is given in Figure~\ref{palo-sen}.  

\subsection{Radial Velocity}\label{secrv}
To derive the planetary mass and confirm the planetary nature of the companion, observations of the reflex motion of the \ko\ parent star were made.  The line-of-sight radial velocity (RV) variations of the parent star were made with the HIRES instrument \cp{Vogt94} on Keck I.  Furthermore, a template spectrum observation was used to determine the stellar \teff, metallicity, and the initial log $g$, using the Spectroscopy Made Easy (SME) tools.  The log $g$ value from spectroscopy was $4.15 \pm 0.05$, considerably lower from the value in the KIC (4.47), but in good agreement with the value obtain from the Markov-Chain Monte Carlo analysis described in \S \ref{secparams}.  The determined \teff\ is $5947\pm100$ K, with a distance estimate of $\sim$600 pc.  $\sim$ We note that the star is chromospherically very quiet.   Our HIRES spectra cover the Ca II H\&K lines, and we measure a chromospheric index, S=0.128 and log R'$_{HK}$ = -5.25, indicating very low magnetic activity, consistent with an old, slowly rotating star.

All but the last four RVs were obtained during the first follow-up season, during the summer of 2009.  The early Keck-HIRES spectra were taken with two compromising attributes.   With a visual magnitude of $V = 13.8$, Kepler-12 was nonetheless observed with short exposure times of typically 10 - 30 minutes, yielding signal-to-noise ratios near SNR=30 per pixel for most spectra.  Such low SNR taxes the Doppler code that was designed for much higher SNR, near 200.  Thus the wavelength scale and the instrumental profile were poorly determined, increasing the RV errors by unknown amounts.   Moreover, all observations except the last four were made with a slit only 2.5 arcsec tall, preventing sky subtraction, which is now commonly applied to HIRES observations of faint \emph{Kepler} stars taken after September 2009.   Moonlight certainly contaminated most of these spectra, as the moon was usually gibbous or full, adding systematic errors to the measured RVs.   Thus the RVs given here contain some poorly known errors that depend on the intensity and Doppler shift of the solar spectra relative to that of the star in the frame of the telescope.  The velocities are given in \mbox{Table \ref{tab:rvs}}.

\begin{figure} \epsscale{0.80}
\plotone{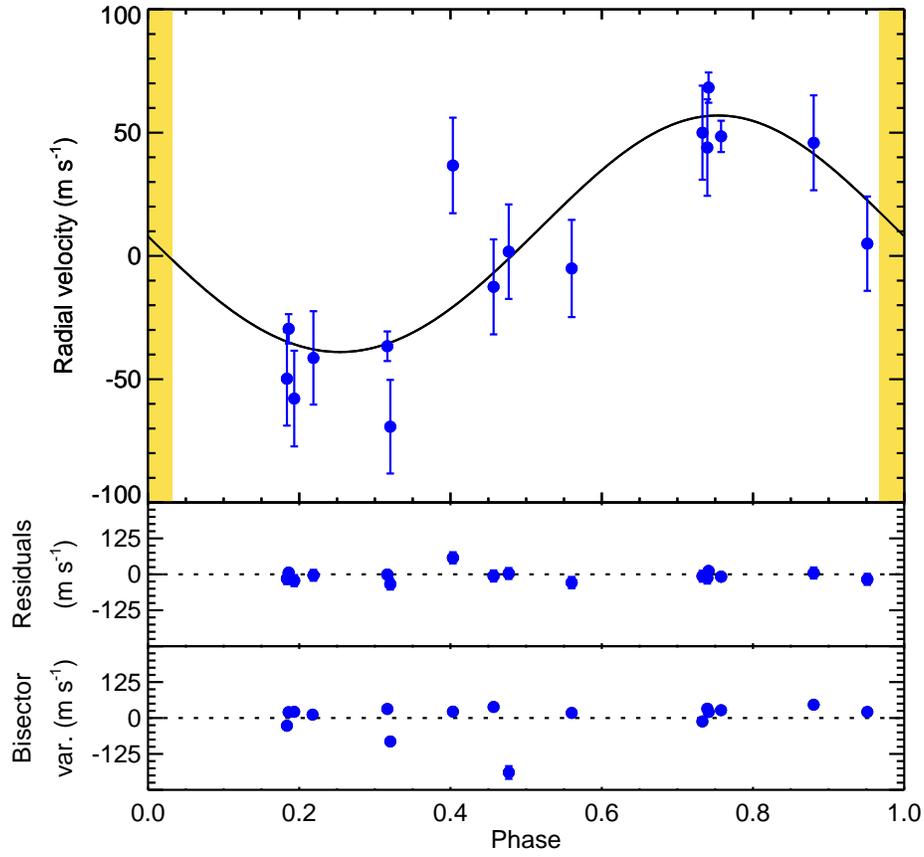}
\caption{(a) Orbital solution for \ko. The observed radial velocities obtained with HIRES on the Keck Telescope are plotted together with the velocity curve for a circular orbit with the period and time of transit fixed by the photometric ephemeris. The radial velocities have an arbitrary zero point. (b) Velocity residuals from the orbital solution. The rms of the velocity residuals is $24~\rm{m\,s^{-1}}$. (c) Variation in the bisector spans for HIRES spectra. The zero point is arbitrary and the RMS is $59~\rm{m\,s^{-1}}$.
\label{rv2}}
\end{figure}

Based on experience with other faint stars similarly observed, we expect true errors close to 18 m s$^{-1}$ due to such effects, which are here included in quadrature.  Orbital analyses should include such uncertainties in applying weights to the RVs, albeit not Gaussian errors.  The largest RV outlier to our orbital analysis is the fourth RV in \mbox{Table \ref{tab:rvs}} and appears at phase 0.4 in Figure \ref{rv2}.  This measurement was made near morning twilight and may be more contaminated than the other measurements by sky spectrum.  However, the measured mass of Kepler-12b is only modestly sensitive to these outliers; the mass of Kepler-12b increases by 7\% when the largest RV outlier to a sinusoidal model is removed and the data are fit again.

The phased radial velocity curve is shown in Figure \ref{rv2}.  Since the orbital ephemeris from \emph{Kepler} photometry was known {\it a priori}, observations were preferentially made at quadrature to allow the most robust determination of planetary mass with the fewest number of RV points.  Observations were also made at additional phases to allow an initial estimate of orbital eccentricity.  The radial velocity observations can be further analyzed for bisector variations, which are shown in Figure \ref{rv2}\emph{c}.  No variation that is in phase with the planetary orbit is found, which supports the planetary nature of the companion.

The radial velocities alone suggest a modest eccentricity, but a circular orbit certainly could not be eliminated with this data set.  Since the long transit duration is the driver towards a large stellar radius, and hence a large planet radius, considerable care was taken to understand if an eccentric orbit around a smaller parent star could lead to the observed transit light curve \cp[e.g.][]{Barnes07b}.  As shown in Sections \ref{secparams} and \ref{2nd}, the timing and duration of the occultation put more robust constraints on eccentricity.

\section{Derivation of Stellar and Planetary Parameters} \label{secparams}
\subsection{Kepler photometry}
Our analysis is based on the Q0-Q7 data, representing nearly 1.5 years of data recorded in a quasi-continuous mode. \textit{Kepler} data are in short- (SC) and long-cadence (LC) timeseries, which are binnings per 58.84876 s and 29.4244 min, respectively, of the same CCD readouts. Eight long-cadence \citep{Jenkins10a} and 16 short cadence \citep{Gilliland10} datasets are used as part of this study, representing 706,135 photometric datapoints and 516 effective days of observations, out of which 464 days have also been recorded in short cadence. We used the raw photometry for our purposes.

\subsection{Data analysis} \label{mcdata}
For this global analysis, we used the implementation of the Markov Chain Monte-Carlo (MCMC) algorithm presented in \citet{Gillon09, Gillon10}. MCMC is a Bayesian inference method based on stochastic simulations that sample the posterior probability distributions of adjusted parameters for a given model. Our MCMC implementation uses the Metropolis-Hasting algorithm \citep[e.g.,][]{Carlin08} to perform this sampling. Our nominal model is based on a star and a transiting planet on a Keplerian orbit about their center of mass.

Our global analysis was performed using 213 lightcurves in total from \textit{Kepler}. For the model fitting we use only the photometry near the transit events. Windows of width 0.8 days (18\% of the orbit) surrounding transits were used to measure the local out-of-transit baseline, while minimizing the computation time. In the analysis 101 SC time-series were used for the transit photometry. The 1-min cadence SC lightcurves yields excellent constraints on the transit parameters \citep[e.g.,][]{Gilliland10,Kipping10}. Furthermore 112 LC time-series were employed for the occultation photometry. Input data to the MCMC also include the 16 RV datapoints obtained from HIRES described in Section \ref{secrv} and the four \textit{Spitzer} 3.6- and 4.5-$\mu$m occultation lightcurves described in Section \ref{sec:spitzer}.

The MCMC had the following set of jump parameters that are  
randomly perturbed at each step of the chains: the planet/star area ratio, the impact parameter $b'=a \cos i /R_{\star}$, the transit duration from first to fourth contact, the time of inferior conjunction $T_0$ (HJD), the orbital period $P$ (assuming no transit timing variations), $K' = K \sqrt{1-e^2} P^{1/3}$, where $K$ is the radial-velocity semi-amplitude, the occultation depth in \textit{Kepler} and both \textit{Spitzer} bandpasses and the two parameters $\sqrt{e}\cos \omega$ and $\sqrt{e}\sin \omega$ \citep{Anderson11}. A uniform prior distribution is assumed for all jump parameters. \textit{Kepler} SC data allow a precise determination of the transit parameters and the stellar limb-darkening (LD) coefficients. We therefore assumed a quadratic law and used $c_1 = 2 u_1 + u_2$ and $c_2 = u_1 - 2 u_2$ as jump parameters, where $u_1$ and $u_2$ are the quadratic coefficients. Those linear combinations help in minimizing correlations on the uncertainties of $u_1$ and $u_2$ \citep{Holman06}.

Three Markov chains of 10$^5$ steps each were performed to derive the system parameters. Their good mixing and convergence were assessed using the Gelman-Rubin statistic \citep{Gelman92}.

\begin{figure} \epsscale{0.80}
\plotone{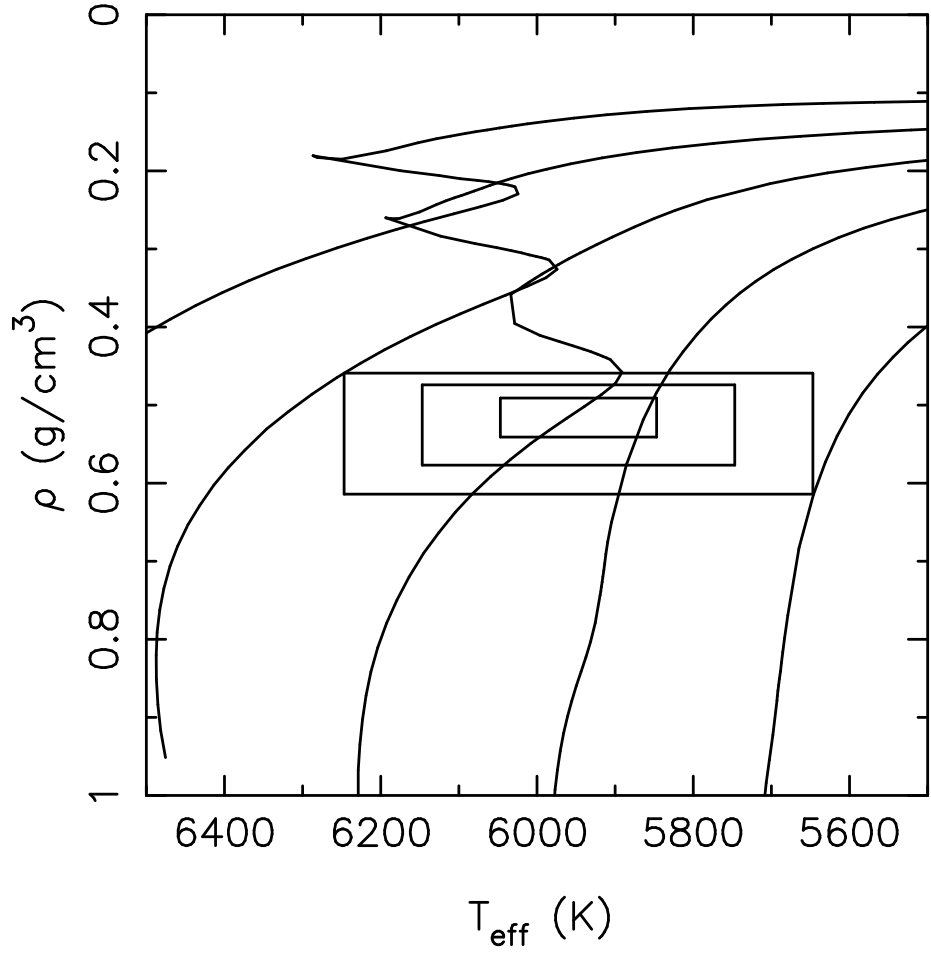}
\caption{Five stellar evolution models from the Yonsei-Yale (Y$^2$) grids.  From left to right the lines show 1.4, 1.3, 1.2, 1.1 and 1.0 \ms\ models for Z=0.0206, which is appropriate given this parent star's metallicity, relative to the solar abundances used in the Y$^2$ grids. The boxes show the 68.3, 95.4, 99.73\% confidence intervals on the stellar \teff\ and \rhostar\ as determined by spectroscopy and transit model fits.  The 'hook' in the evolution tracks for more massive stars with convective cores produces a non-uniform distribution of masses with the uncertainty boxes and can produce a degeneracy in the solution for the best-fit stellar parameters.
\label{track}}
\end{figure}

At each step, the physical parameters are determined from the jump parameters above and the stellar mass.  The transit and radial velocity measurements together determine the planet orbit and allow for a geometrical measure of the mean density of the host star (\rhostar). Using the MCMC chains, the probability distribution on \rhostar\ was calculated and together with the spectroscopically measured values and uncertainties of \teff\ and \feh\ are used to determine consistent stellar parameters from Yonsei-Yale stellar evolution models \citep{Demarque04}.  The derived stellar \teff\ and \rhostar\ parameters, compared to stellar evolution tracks, are shown in Figure \ref{track}.  The resulting normal distribution aroud the stellar mass ($1.166\pm0.052$) \ms\ was then used as a prior distribution in a new MCMC analysis, allowing the physical parameters of the system to be derived at each step of the chains.

\subsubsection{Model and systematics}
The \textit{Kepler} transit and occultation photometry are modeled with the \citet{Mandel02} model, multiplied by a second order polynomial accounting for stellar and instrumental variability. We added a quadratic function of the PSF position to this baseline model for the \textit{Spitzer} occultation lightcurves (see Section \ref{sec:spitzer}).

Baseline model coefficients are determined for each lightcurve with the Singular Value Decomposition (SVD) method \citep{Press92} at each step of the MCMC. 
Correlated noise was accounted for following \citet{Winn08,Gillon10}, to ensure reliable error bars on the fitted parameters. For this purpose, we computed a scaling factor based on the standard deviation of the binned residuals for each lightcurve with different time bins. The error bars are then multiplied by this scaling factor. We obtained a mean scaling factor of 1.02 for all \textit{Kepler} photometry, denoting a negligible contribution from correlated noise. The mean global \textit{Kepler} photometric RMS per 30-min bin is 159 parts per million (ppm).

\subsection{Results}
\begin{figure}
\epsscale{1.0}
\plotone{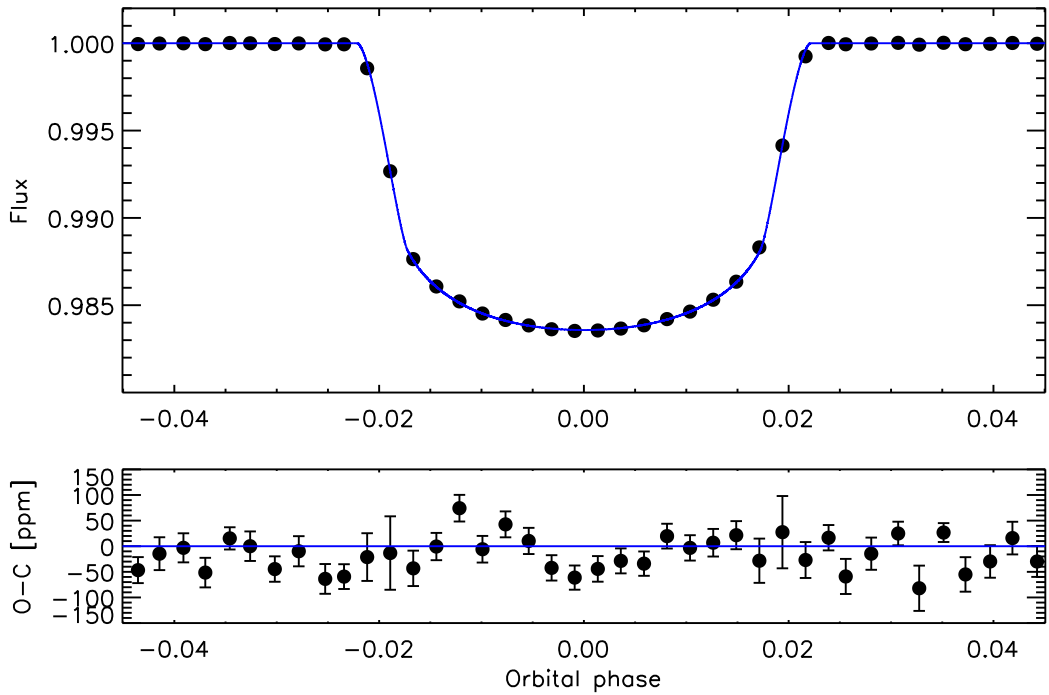}
\caption{Top: Kepler-12\,b phase-folded transit lightcurve with best-fit model superimposed. The data are binned in 15 minute intervals. Error bars are smaller than the plotted datapoints. Bottom: residuals are displayed in parts per million (ppm) scale. \label{fig1}}
\end{figure}

We show in \mbox{Table \ref{tab:params}} the median values and the corresponding 68.3\% probability interval of the posterior distribution function (PDF) for each parameter obtained from the MCMC. We present in Figure \ref{fig1} the phase-folded transit photometry.  We determine a planetary radius of $1.695^{+0.028}_{-0.032}$ \rj\ and a mass of $0.431^{+0.041}_{-0.040}$ \mj\ that produces a very low mean planetary density of $0.111^{+0.011}_{-0.010}$ g\,cm$^{-3}$.

We measure occultation depths of $0.099\pm0.028$\% and $0.116\pm0.034$\% in \textit{Spitzer} IRAC 3.6 and 4.5$\mu$m channels respectively, consistent at the 1$\sigma$ level with the specific analysis present in Section \ref{sec:spitzer}.  The LD quadratic coefficients derived from the MCMC are $u_1 = 0.375 \pm 0.004$ and $u_2 = 0.250 \pm 0.008$. Those are in good agreement with the theoretical coefficients obtained from the \citet{Claret2011} tables of $u_1 = 0.366$ and $u_2 = 0.275$.

We finally determine an occultation depth of $31\pm$8 ppm in the \textit{Kepler} bandpass, which corresponds to a geometric albedo $A_g=0.14\pm0.04$.  The geometric albedo is wavelength-dependent and measures the ratio of the planet flux at zero phase angle to the flux from a Lambert sphere at the same distance and the same cross-sectional area as the planet \cp[see, e.g.,][]{Marley99,Sudar00}:
\begin{equation}
\frac{F_p}{F_\star}=A_g\left( \frac{R_p}{a}\right)^2
\end{equation}
where $\frac{F_p}{F_\star}$ is the occultation depth, $a$ the orbital semi-major axis and $R_p$ the planetary radius.

\begin{figure}
\epsscale{1.0}
\plotone{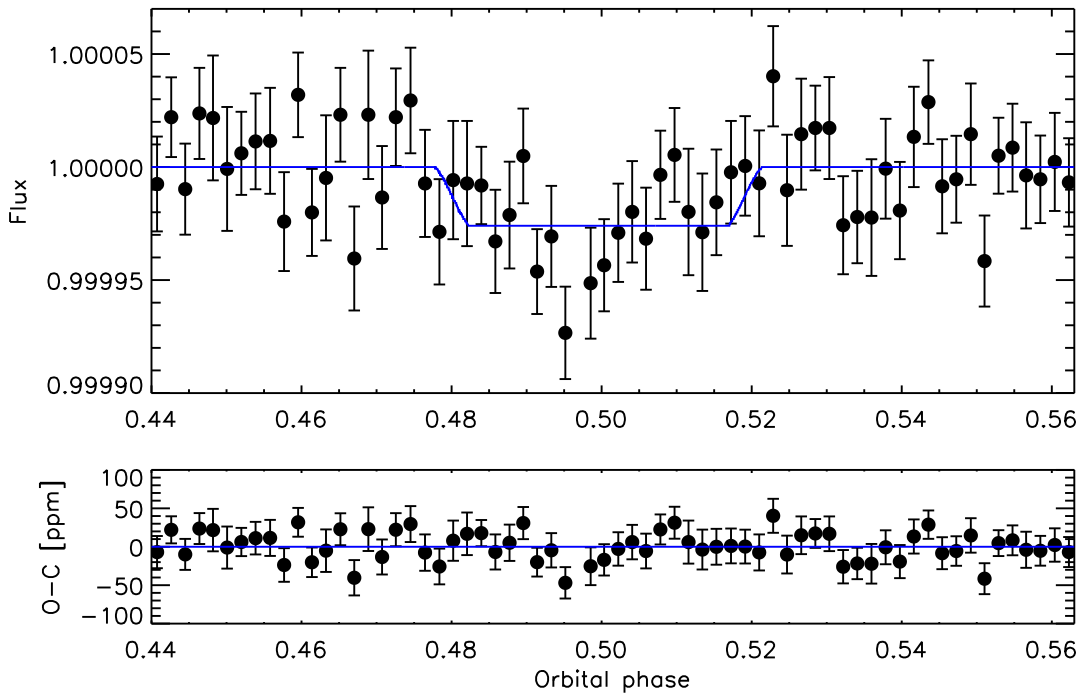}
\caption{Top: Kepler-12\,b phase-folded occultation lightcurve with best-fit model superimposed.  The \emph{Kepler} data are binned in 15 minute increments. Bottom: residuals are displayed in parts per million (ppm) scale.\label{fig2}}
\end{figure}

\begin{figure}
\epsscale{1.0}
\plotone{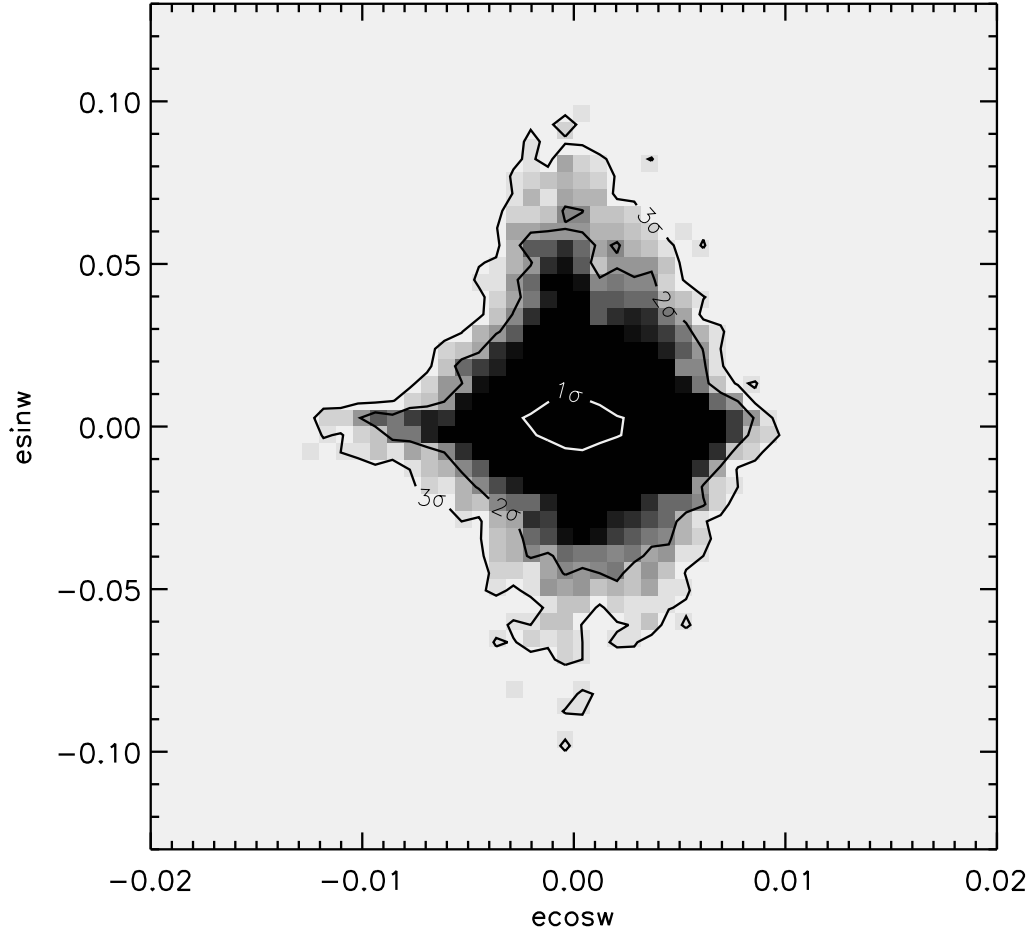}
\caption{Density function of the two-dimensional $e \sin \omega$/$e \cos \omega$ successful MCMC trials (density increases from white to black).
Note the different scales for x and y axes. The 1, 2 and 3$\sigma$ confidence domains are superimposed. The weak constraint from the 
occultation duration leads to a significant scatter in $e \sin \omega$. As $e \cos \omega$ is well constrained from the occultation timing obtained by the \emph{Kepler} and \emph{Spitzer} photometry,
this yields orbital eccentricities as high as $\sim~0.09$ (3$\sigma$ upper limit) provided the argument of periastron is close to 90 or 270 degrees.
In this case, the eccentricity vector points toward or away of the observer, allowing a wide range of $e \sin \omega$ values while leaving $e \cos \omega$ almost unchanged.\label{fig3}}
\end{figure}

The corresponding phase-folded occultation lightcurve is shown in Figure \ref{fig2}.  The combination of \textit{Spitzer} and \textit{Kepler} occultations leads to a 1$\sigma$ orbital eccentricity signal of $e<0.01$, while the 3$\sigma$ limit is $e<0.09$. We show $e \sin \omega$ vs.~$e \cos \omega$ from successful MCMC trials in Figure \ref{fig3}.  The small allowed eccentricity removes most solutions that allow fits to the long transit duration with smaller stellar (and planetary) radii.  There are two paths towards a more robust constraint on $e$.  One would come from many additional RV points.  An easier path would be additional quarters of \emph{Kepler} data, which would yield a better determination of the occultation duration, which constrains $e \sin \omega$.  All system parameters are collected in \mbox{Table \ref{tab:params}}.

\section{Atmospheric Characterization at Secondary Eclipse} \label{2nd}
As part of \emph{Spitzer} program \#60028 (D.~Charbonneau, PI) a number of {\it Kepler}-detected giant planets were observed in order to characterize the planets' thermal emission at 3.6 and 4.5 $\mu$m  during the Warm \emph{Spitzer} extended mission.  The inherent faintness of the planetary targets mean some stars must be observed more than once for adequate signal-to-noise to enable meaningful atmospheric characterization.

In addition to the measurement of the depth of the occultation (or secondary eclipse), which yields a measurement of the planetary brightness temperature, the timing and duration of the occultation constrains $e$, as described above.  The timing of the transit constrains $e \cos \omega$ where $\omega$ is the longitude of periapse.  The duration of the transit constrains $e \sin \omega$.  The former is generally easier to measure accurately than the latter.

\subsection{Warm Spitzer Detections}\label{sec:spitzer}

Kepler-12 was observed during four occultations between August 2010 and January 2011 with \wspitzer/IRAC \citep{werner04,fazio04} at 3.6 and 4.5~\micron. Two occultations were gathered per bandpass and each visit lasted approximately 11~h. 
The data were obtained in full-frame mode ($256\times256$ pixels) with an exposure time of
30.0~s per image which yielded 1321 images per visit. 
The set of observations are shown in \mbox{Table~\ref{tab:spitzer}}.

The method we used to produce photometric time series from the images is described in \cite{desert11b}.
It consists of finding the centroid position of the stellar point spread function (PSF) and performing aperture photometry using a circular aperture  on individual exposures.
The images used are the Basic Calibrated Data (BCD) delivered by the \emph{Spitzer} archive.
These files are corrected for dark current, flat-fielding, detector non-linearity and converted into flux units.  
We convert the pixel intensities to electrons using the information on detector gain and exposure time provided in the FITS headers.
This facilitates the evaluation of the photometric errors.
We extract the UTC-based Julian date for each image from the FITS header and correct to mid-exposure. 
We then correct for transient pixels in each individual image using a 20-point sliding median filter of the pixel intensity versus time.
To do so, we compare each pixel's intensity to the median of the 10 preceding and 10 following exposures at the same pixel position and we replace outliers greater than $3~\sigma$ with its median value.
The fraction of pixels we correct varies between 0.15\% and 0.22\% depending on the visit.
The centroid position of the stellar PSF is determined using DAOPHOT-type Photometry Procedures, \texttt{GCNTRD}, from the IDL Astronomy Library\footnote{{\tt http://idlastro.gsfc.nasa.gov/homepage.html}}. We use the \texttt{APER} routine to perform aperture photometry with a circular aperture of variable radius, using radii of $1.5$ to $8$ pixels, in $0.5$ steps. 
The propagated uncertainties are derived as a function of the aperture radius; we adopt the one which provides the smallest errors. 
We find that the transit depths and errors vary only weakly with the aperture radius for all the light-curves analyzed in this project. 
The optimal apertures are found to have radii of $2.5$~pixels. 

\begin{figure*}[h!]
\begin{center}
 \includegraphics[width=4.5in]{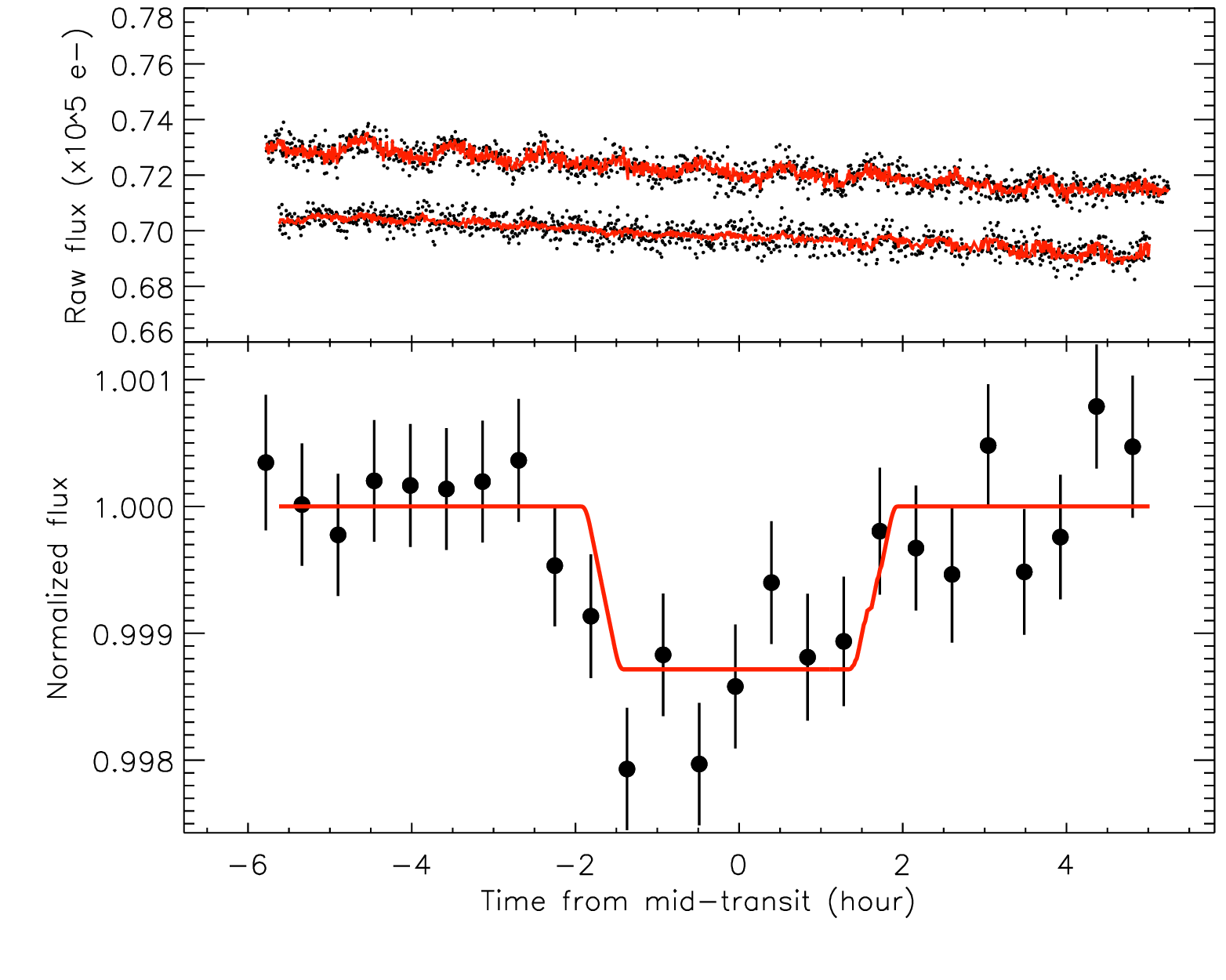}
 \includegraphics[width=4.5in]{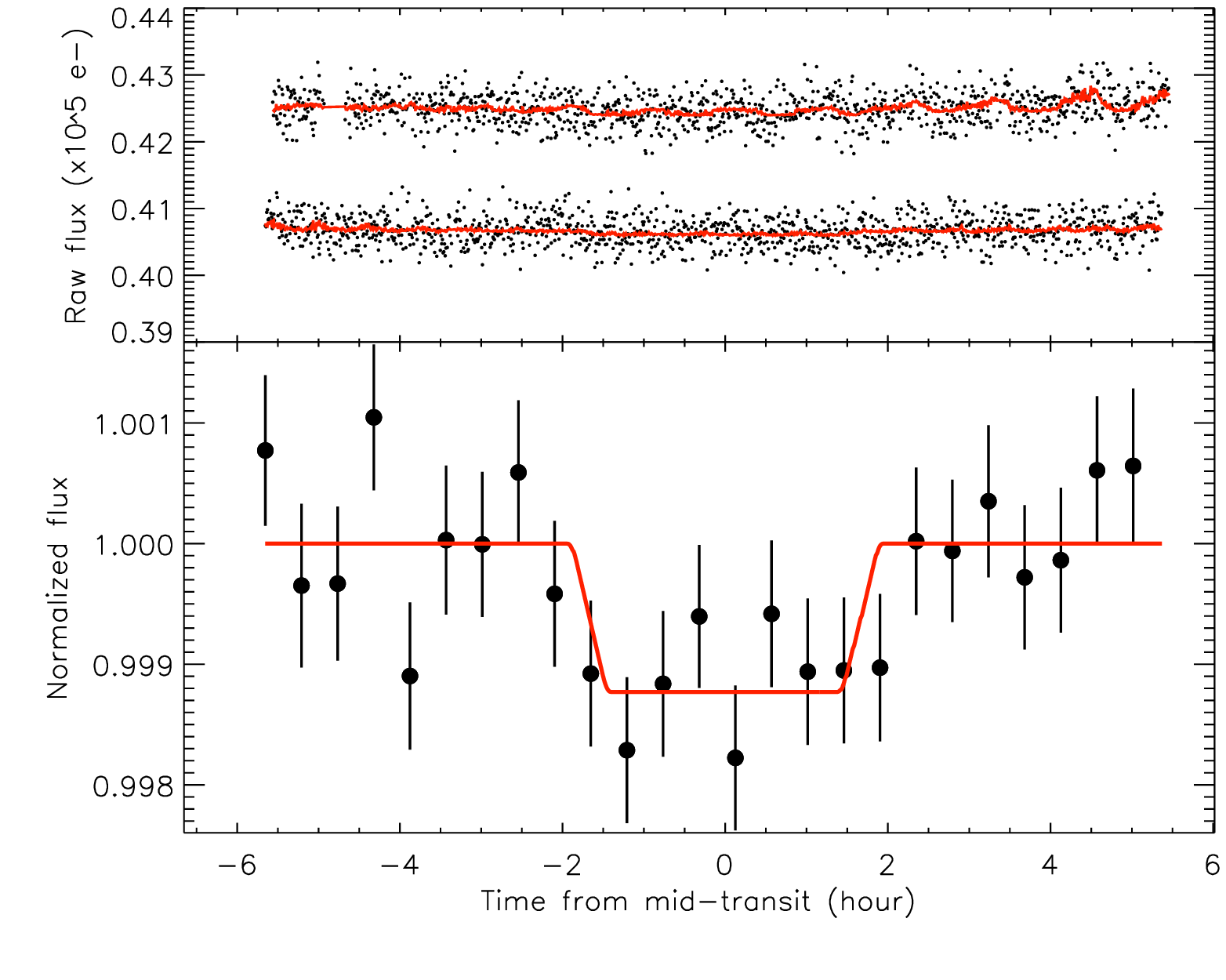}
 \caption{\spitzer\ occultation light-curves of \koi\ observed in the IRAC band-pass at 3.6 (top) and 4.5~\micron\ (bottom). Top panels: raw and unbinned light-curves. The red solid lines correspond to the best fit models which include the time and position instrumental decorrelations as well as the model for the planetary transit (see details in Sect.~\ref{sec:spitzer}). Bottom panels: corrected and normalized occultation light-curve with the best fit model (in red). The data are binned in 25~minutes intervals (50 points).}
   \label{fig:spitzerlightcurves}
\end{center}
\end{figure*}

We estimate the background by fitting a Gaussian to the central region of the histogram of counts from the full array. 
The center of the Gaussian fit is adopted as the residual background intensity.
As already seen in previous \wspitzer\ observations \citep{deming11,beerer11}, we find that the background varies by 20\% between three distinct levels from image to image, and displays a ramp-like behavior as function of time.
The contribution of the background to the total flux from the stars is low for both observations, from 0.07\% to 1.2\% depending on the images. 
Therefore, photometric errors are not dominated by fluctuations in the background. 
We used a sliding median filter to select and trim outliers in flux and positions greater than $4~\sigma$.
This process removes between $0.9\%$ and $2.8\%$ of the data, depending on the visit. 
We also discarded the first half-hour of observations, which are affected by a significant telescope jitter before stabilization. 
The final number of photometric measurements used are presented in \mbox{Table~\ref{tab:spitzer}}. 
The raw time series are presented in the top panels of Figure~\ref{fig:spitzerlightcurves}.

We find that the point-to-point scatter in the photometry gives a typical signal-to-noise ratio of $260$ and $200$ per image  at 3.6 and 4.5~\micron\ respectively. These correspond to 85\% of the theoretical signal-to-noise.
Therefore, the noise is dominated by Poisson photon noise.
We used a transit light curve model multiplied by instrumental
decorrelation functions to measure the occultation parameters and their
uncertainties from the \spitzer\ data as described in \cite{desert11a}. 
We compute the transit light curves with the IDL transit routine
\texttt{OCCULTSMALL} from \cite{Mandel02}.
In the present case, this function depends on one
parameter: the occultation depth $d$.
The planet-to-star radius ratio $R_p / R_\star$, the orbital semi-major axis to stellar radius ratio (system scale) $a / R_\star$, the mid-occultation time $T_c$ and the impact parameter $b$ are set fixed to the values derived from the \kepler\ lightcurves.

The \spitzer/IRAC photometry is known to be systematically affected by the
so-called \textit{pixel-phase effect} (see e.g., \citealt{Charb05,Knutson08}).
This effect is seen as oscillations in the measured fluxes with a period of approximately 70~min (period of the telescope pointing jitter) and an amplitude of approximately $2\%$ peak-to-peak. 
We decorrelated our signal in each channel using a linear function of time for the baseline (two parameters) and a quadratic function of the PSF position (four parameters) to correct the data for each channel.
We performed a simultaneous Levenberg-Marquardt least-squares fit \citep{markwardt09} to the
data to determine the occultation depth and instrumental model parameters (7 in total).
The errors on each photometric point were assumed to be identical, and were set to the $rms$ of the residuals of the initial best-fit model.
To obtain an estimate of the correlated and systematic errors \citep{pont06} in our measurements, we use the residual permutation bootstrap, or ``Prayer Bead'', method as described in \citet{desert09}. In this method, the residuals of the initial fit are shifted systematically and sequentially by one frame, and then added to the transit light curve model
before fitting again. 
We allow asymmetric error bars spanning $34\%$ of the points above and below the median of the distributions to derive the $1\sigma$ uncertainties for each parameter as described in \citet{desert11b}.

We measure the occultation depths in each bandpass and for each individual visit. 
The values we measure for the depths are all in agreement at the 1$\sigma$ level.
Furthermore the weighted mean averages per bandpass of the transit depths are consistent with the depths derived by the global Monte-Carlo analysis.

\subsection{Joint Constraints on the Atmosphere}

To model the planet's atmosphere we use a one-dimensional plane-parallel atmosphere code that has been widely used for solar system planets, exoplanets, and brown dwarfs over the past two decades.  The optical and thermal infrared radiative transfer solvers are described in detail in \ct{Toon89}.  Past applications of the model include Titan \cp{Mckay89}, Uranus \cp{MM99}, gas giant exoplanets \cp{Fortney06,Fortney07b,Fortney08a}, and brown dwarfs \cp{Marley96,Burrows97,Marley02,Saumon08}.  We use the correlated-k method for opacity tabulation \cp{Goody89}.  Our extensive opacity database is described in \ct{Freedman08}.  We make use of tabulations of chemical mixing ratios from equilibrium chemistry calculations of K.~Lodders and collaborators \cp{Lodders99,Lodders02,Lodders06}.  We use the protosolar abundances of \ct{Lodders03}.  Since the first detection of thermal flux from hot Jupiters \cp{Charb05,Deming05b} we have used the code extensively to model strongly irradiated planet atmospheres and have compared model spectra to observations \cp[e.g.][]{Fortney05,Knutson08b,deming11,desert11b}.

\begin{figure}\epsscale{0.8}
\plotone{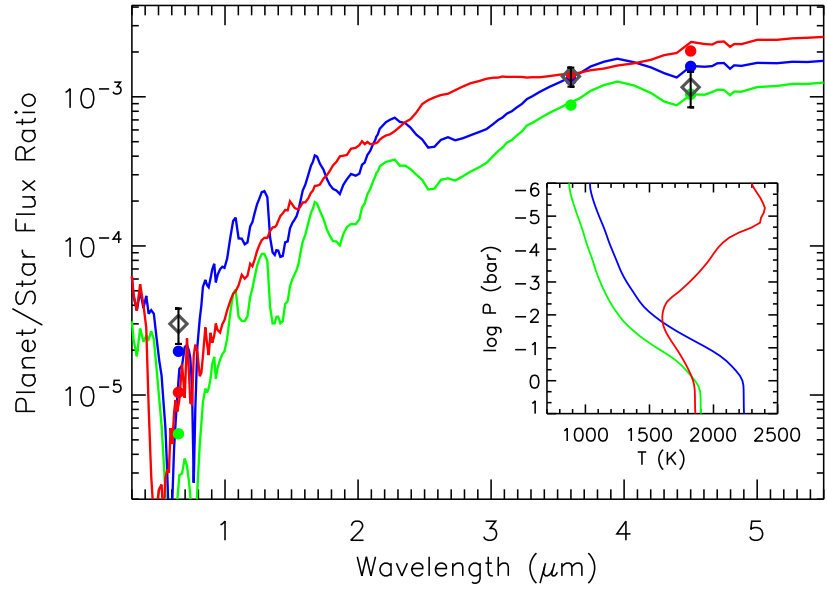}
\caption{\emph{Main Panel}: Planet-to-star flux ratios observed by \kk\ and \emph{Spitzer}, shown in gray.  The flux ratios from three planetary models are shown for comparison.  There is preference towards models that have no temperature inversion (blue and green).  Model ratios integrated over the appropriate bandpasses are shown as filled circles.  The \kk\ occultation point strongly argues for inefficient redistribution of flux, or an additional scattering component at optical wavelengths. \emph{Inset Panel}:  Atmospheric pressure-temperature profiles for the three models.  
\label{ratios}}
\end{figure}

Planet \ko\ intercepts an incident flux of $1.1 \times 10^9$ erg s$^{-1}$ cm$^{-2}$, a value just larger than the suggeste pM/pL class incident flux boundary proposed by \ct{Fortney08a}.  It was suggested that planets warmer than this boundary (pM) would harbor dayside temperature inversions, while those cooler than this boundary would not have inversions.  It is therefore important to understand the temperature structure of the planet.  For \ko\ we show three models in Figure \ref{ratios}, for which we plot the planet-to-star flux ratio and dayside \emph{P--T} profiles. In red and blue are ``dayside average'' models with incident flux redistributed over the dayside only.  In green is a model where the incident flux is cut in half, to simulate efficient redistribution of energy to the night side \cp[see, e.g.][]{Fortney07b}. The model in red has a temperature inversion due to absorption of incident flux by TiO and VO vapor \cp[e.g.][]{Hubeny03,Fortney08a}, while the blue and green models lack inversions, as TiO/VO vapor is removed from the opacities.  The \kk\ occultation depth is shown at 0.65 $\mu$m (diamond), while the \emph{Spitzer} detections are shown as diamonds at 3.6 and 4.5 $\mu$m.  Model band-averages at these wavelengths are shown as solid circles.

The relatively flat ratio of the 3.6/4.5 diamond points generally points to a very weak or no inversion \cp{Knutson10}.  Looking to the optical, the green model is dramatically too dim, while the blue model nearly reaches the 1$\sigma$ error bar.  Looking at the infrared, the blue point is at the 1$\sigma$ 4.5 $\mu$m error bar as well.  The inverted model (red) has approximately the same \teff\ ($\sim~1700$ K) as the blue model, but higher fluxes in the mid infrared and lower fluxes in the near-infrared and optical.  The \emph{Spitzer} data alone do not give us strong leverage on the temperature structure.  Any cooler model with an inversion (not plotted) would yield a better fit to \emph{Spitzer} and a worse fit to \kk.  Within the selection of models, the brightness of the \kk\ point argues for the no-inversion model.  The flux in the \kk\ band from the blue model is 60\% scattered light, 40\% thermal emission.

Our tentative conclusion is that the blue (no inversion, inefficient temperature homogenization onto the night side) model is preferred.  However, given our ignorance of the optical opacity in these atmospheres, this conclusion is tentative.  The relatively deep occultation in the \kk\ band argues for an additional contribution at optical wavelengths that is not captured in the model.  One possibility is that stellar flux has photoionized Na and K gasses \cp{Fortney03}, which are thought to be strong absorbers of stellar light (and therefore diminish scattered light) in hot Jupiter atmospheres.  Another possibility is a population of small grains, such as silicates, which could scatter some stellar flux \cp{Marley99,SWS00,Sudar00}.  Such clouds are prominent in L-dwarf atmospheres \cp[e.g.][]{AM01}.

\section{Discussion} \label{discuss}
A great number of explanations have been put forward to explain the inflated radii of the close-in giant planets.  They generally fall into several broad classes, and are recently reviewed in \ct{Fortney10} and \ct{Baraffe10}.  Some argue for a delayed contraction, due to slowed energy transport in the atmosphere \cp{Burrows07} or the deep interior \cp{Chabrier07c}.  Others suggest a variety of atmospheric affects \cp{Showman02,Guillot02,Batygin10,Arras10,Youdin10} that lead to energy dissipation into the interior.  Still others suggest tidal dissipation in the interior due to eccentricity damping \cp{Bodenheimer00,Jackson08,Miller09,Ibgui09}.  

For \koi\ we do not find evidence for transit timing variations \cp{Ford11}.  The RMS scatter of transit times about a
linear ephemeris is less than one minute and only 17\% larger than the
average of the formal timing uncertainties.  This rules out the
presence of  massive non-transiting planets very near by or in the
outer 1:2 mean motion resonance.  In principle, a more distant
non-resonant planet is possible, but hot Jupiters rarely have a second
massive planet close to the star \cp{Wright09,Wright11,Latham11}.  Thus, it
is very unlikely that the inflated radius is due to eccentricity
damping.

Clarity on a radius-inflation mechanism has not been achieved, but Figure \ref{rvsi} appears to argue for an explanation based on the planet temperature or irradiation level of the atmosphere (rather than merely on orbital separation), as has been shown by other authors \cp{Kovacs10,Laughlin11,Batygin11}.

If the inflation mechanism can be thought of as an energy source that is added to the planet's deep convective interior, we can readily compare the energy input needed to sustain the radius of \ko, compared to other planets.  This is actually more physically motivated than the more commonly discussed ``radius anomaly,'' since the power needed to inflate the radius by a given amount, $\Delta R$, is a very strong function of mass.   In particular, Figure 6 in \ct{Miller09} allows for a comparison of input power as a function of planet mass, for 4.5 Gyr-old model planets with 10 \me\ cores at 0.05 AU from the Sun.  For instance, inflating a 0.2 \mj\ planet by 0.2 \rj\ over its expected radius value takes $1 \times 10^{24}$ erg s$^{-1}$, while for a 2 \mj\ planet it is $2 \times 10^{27}$ erg $^{-1}$, a factor of 2000 difference in power for a factor of 10 in mass.  This is the reason why \ct{Batygin11} can easily expand Saturn-mass planets to the point of disruption via Ohmic dissipation---a small amount of energy goes a long way towards inflating the radii of low-mass planets. 

In understanding the structure of \ko, we can use the models described in \ct{Miller09}, which are adapted from \ct{Fortney07a}.  In particular, Table 1 in \ct{Miller09} includes the current internal power necessary to explain the radius of several inflated planets.  Planets \hd\ \cp{Henry00,Charb00} and TrES-4b \cp{Mandushev07} are interesting points of comparison.  \hd\ and \ko\ have similar incident fluxes, while TrES-4b and \ko\ have similar inflated radii.

Since \ko\ and \hd\ have comparable incident stellar fluxes (that of \ko\ is 14\% larger), one could easily assume that they have similar interior energy sources \cp[e.g.][]{Guillot02}.  For \hd, with core masses of 0, 10, and 30 \me, incident powers of $1.5\times10^{26}$, $3.8\times10^{26}$, and $1.6\times10^{27}$ erg s$^{-1}$, are required \cp{Miller09}.  Using the planetary parameters of \ko\ with cores of 0, 10, and 30 \me, the required powers are substantially larger.  The enhancement is generally a factor of three larger, with values of $4.4\times10^{26}$, $1.1\times10^{27}$, and $4.2\times10^{27}$ erg s$^{-1}$, respectively.  This could point to more than one radius inflation mechanism being at play in this planet, as has recently been strongly suggested for the massive transiting planet CoRoT-2b \cp{Guillot11}.

\begin{figure}
\plotone{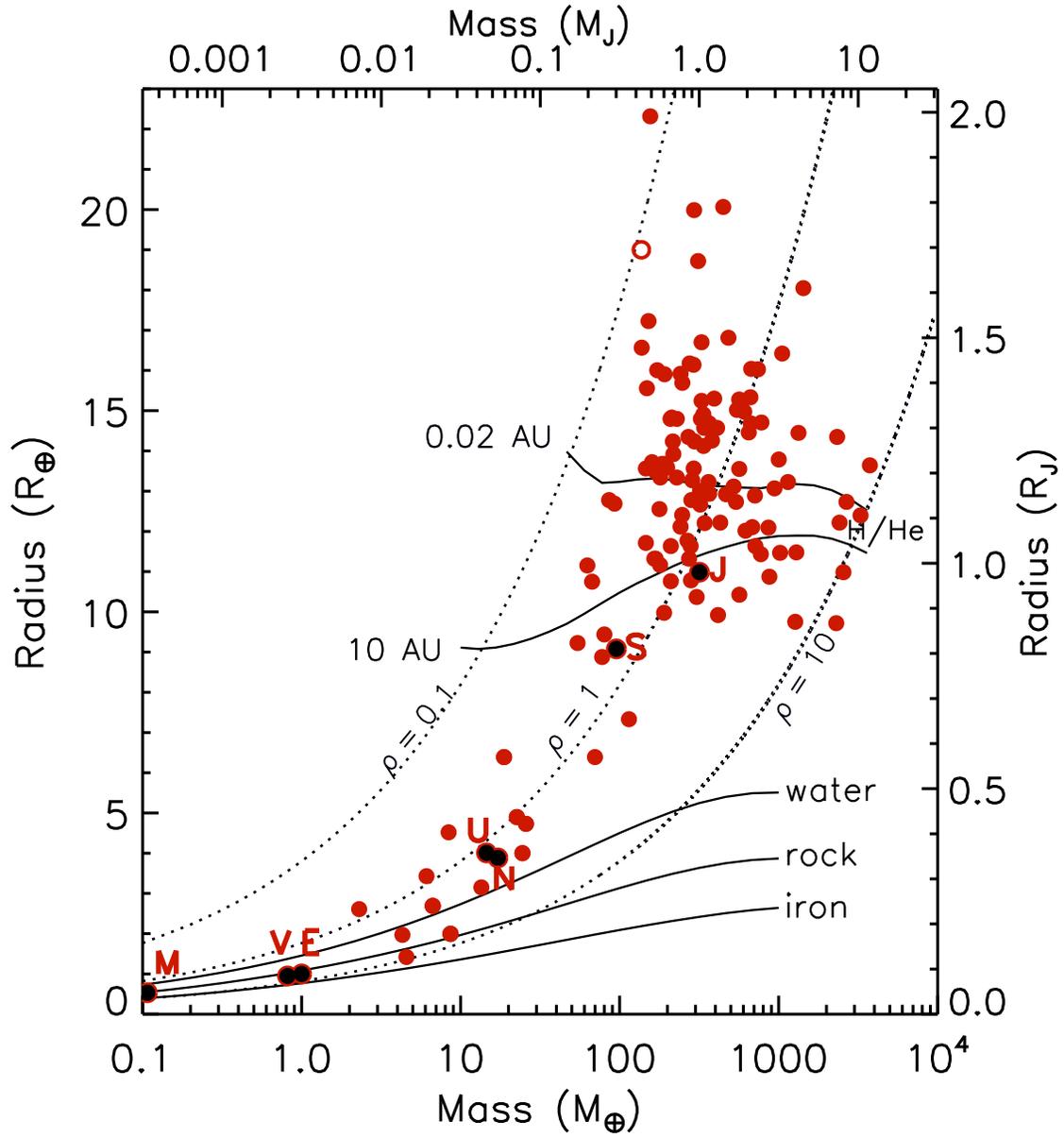}
\caption{Mass vs.~radius for planets with ``well-defined parameters," as taken from  http://www.inscience.ch/transits/, but also including the Kepler-11 system \cp{Lissauer11}.  \ko\ is shown as an open circle, the 2nd-lowest density planet discovered.  Models (solid black curves) are taken from \ct{Fortney07a}.  The two upper curves are for pure H-He planets, at 4.5 Gyr, at 0.02 and 10 AU from the Sun.
\label{mrplot}}
\end{figure}

The difference between \ko\ and \hd\ can be remedied, however, if the planets have different heavy element masses.  In particular, both planets would require power levels of $\sim 1.6 \times10^{27}$ erg s$^{-1}$ if \ko\ possesses only $\sim$~15 \me\ of heavy elements, while \hd\ possesses 30 \me.  The \ko\ parent star has an [Fe/H]=+0.07, while for HD 209458 it is +0.02.  As recently shown by \citet{Miller11} for the colder non-inflated planets, for parent stars with similar stellar metallicities, a spread from 10-30 \me\ is reasonable.  Therefore it appears that the wide disparity in radii between these two well-studied planets could alternatively be due to the differences in interior heavy element masses.  Large diversities in heavy element abundances are clearly needed to explain plots like Figure \ref{mrplot}, where planets of similar masses can have dramatically different radii.

For comparison, TrES-4b at 0.93 \mj\ and 1.78 \rj\ is nearly twice as massive as \ko, but intercepts 2.1 times higher incident flux.  The inflation powers at 0, 10, and 30 \me\ range from 1.0 to $3.4\times10^{28}$, 8-20 times larger than for \ko, at the same heavy element masses.  Clearly the required energy difference between the two models does not scale simply with the incident flux.  As discussed in \ct{Miller11} as the population of cool (\teq $<1000$ K) non-inflated planets grows, the heavy element mass of extrasolar gas giants can become better understood as a function of planet mass and stellar metallicity, which will allow for more robust constraints on the heavy element masses of the inflated planets.  This will in turn allow for better estimates of the magnitude of the additional interior energy source within these planets.  While \ko\ does not quite fit the general trend that the highest irradiation planets are the largest, this trend argues for an explanation that scales with atmospheric temperature.  A more detailed computational understanding of how the visible atmosphere, deep atmosphere, and convective interior interact and feedback on each other is now clearly needed.

\section{Conclusions} \label{conclu}
We report the discovery of planet \ko\ from transit observations by \emph{Kepler}.  The planet has an unusually inflated radius and low bulk density.  At its incident flux level, the large radius of the planet makes it somewhat of an outlier compared to the general empirical trend that the most inflated planets intercept the highest incident fluxes.  This may require the planet to have an usually low mass fraction of heavy elements within its interior, or that more than one radius-inflation mechanism is at work in its interior.

The atmosphere of the planet is probed via detections of the occultations in the \emph{Kepler} and Warm \emph{Spitzer} bandpasses.  Given the faintness of the parent star, characterization was difficult, but all detections were made at a level of at least 3.5$\sigma$.  A model comparison to the data yields a best-fit model that lacks a dayside temperature inversion, given the relatively flat 3.6/4.5 $\mu$m ratio of the planet-to-star flux ratios, along with the relatively large occultation depth in the \emph{Kepler} band.  Additional \kk\ data will yield more robust constraints on the planet's geometric albedo, orbital eccentricity, and perhaps phase curve information.

\acknowledgements
JJF acknowledges the support of the Kepler Participating Scientist's program, via NASA grant NNX09AC22G.  \emph{Kepler} was competitively selected as the tenth Discovery mission.  Funding for the \kk\ mission is provided by NASA's Science Mission Directorate. We thank the many people who gave so generously of their time to make the \kk\ mission a success.  We thank Carly Chubak for cross correlation analyses of the Keck spectra for companions.  This work incorporates observations made with the \emph{Spitzer Space Telescope}, which is operated by the Jet Propulsion Laboratory, California Institute of Technology under a contract with NASA. Support for this work was provided by NASA through an award issued by JPL/Caltech.  Finally, the authors wish to extend special thanks to those of Hawai\`ian ancestry on whose sacred mountain of Mauna Kea we are privileged to be guests. Without their generous hospitality, the Keck observations presented herein would not have been possible.


\begin{deluxetable}{lrrrr}
\tablewidth{0pc}
\tablecaption{
	Relative radial velocity and bisector span variation measurements of Kepler-12.
	\label{tab:rvs}
}
\tablehead{
	\colhead{BJD} &
	\colhead{RV} & 
	\colhead{\ensuremath{\sigma_{\rm RV}}} &
	\colhead{BS} &
	\colhead{\ensuremath{\sigma_{\rm BS}}} \\
	\colhead{ } &
	\colhead{($\rm{m~s^{-1}}$)} &
	\colhead{($\rm{m~s^{-1}}$)} &
	\colhead{($\rm{m~s^{-1}}$)} &
 \colhead{($\rm{m~s^{-1}}$)} 
}
\startdata
$ 2455014.91234 $ & $  -12.5 $ & $ 19.3 $ & $ 38.3 $ & $ 6.1 $ \\
$ 2455016.79104 $ & $  45.9 $ & $ 19.3 $ & $ 46.2 $ & $ 6.7 $ \\
$ 2455017.10568 $ & $  5.0 $ & $ 19.1 $ & $ 21.1 $ & $ 3.4 $ \\
$ 2455019.11155 $ & $  36.7 $ & $ 19.4 $ & $ 22.3 $ & $ 4.8 $ \\
$ 2455027.05631 $ & $  -57.9 $ & $ 19.4 $ & $ 21.5 $ & $ 4.6 $ \\
$ 2455041.99855 $ & $  -5.1 $ & $ 19.7 $ & $ 17.6 $ & $ 5.9 $ \\
$ 2455042.79449 $ & $  44.0 $ & $ 19.6 $ & $ 32.3 $ & $ 4.6 $ \\
$ 2455073.83165 $ & $  50.0 $ & $ 19.1 $ & $ -12.1 $ & $ 2.9 $ \\
$ 2455075.83091 $ & $  -49.8 $ & $ 19.0 $ & $ -27.0 $ & $ 2.9 $ \\
$ 2455080.87564 $ & $  -69.3 $ & $ 19.0 $ & $ -81.8 $ & $ 9.1 $ \\
$ 2455084.86254 $ & $  -41.4 $ & $ 18.9 $ & $ 11.7 $ & $ 4.4 $ \\
$ 2455134.82606 $ & $  1.7 $ & $ 19.2 $ & $ -189.6 $ & $ 22.1 $ \\
$ 2455437.78012 $ & $  68.3 $ & $ 6.1 $ & $ 21.0 $ & $ 3.5 $ \\
$ 2455439.75382 $ & $  -29.6 $ & $ 6.0 $ & $ 20.2 $ & $ 3.7 $ \\
$ 2455759.86617 $ & $  -36.6 $ & $ 6.0 $ & $ 31.4 $ & $ 3.8 $ \\
$ 2455761.82490 $ & $  48.5 $ & $ 6.4 $ & $ 27.1 $ & $ 4.1 $ 
\enddata
\end{deluxetable}                                                       


\begin{deluxetable}{ll}
\tablecaption{Kepler-12 system parameters}
\tablehead{\colhead{Parameters} & \colhead{Value} }
\startdata
\textit{Jump parameters} &   \\
\tableline
 &   \\
Planet/star area ratio $(R_p/R_s)^2$ & $0.013765^{+0.000020}_{-0.000020}$ \\
$b'=a \cos i /R_{\star}$ [$R_{\star}$] & $0.174^{+0.011}_{-0.011}$ \\
Transit width [d] & $0.19573^{+0.00009}_{-0.00010}$ \\
$T_0$ - 2450000 [HJD] & $5004.00835^{+0.00002}_{-0.00002}$ \\
Orbital period $P$ [d] & $4.4379637^{+0.0000002}_{-0.0000002}$  \\
RV $K'$ [m\,s$^{-1}$\,d$^{1/3}$]  & $79.2^{+7.1}_{-7.0}$  \\
$\sqrt{e} \cos \omega$ & $-0.001^{+0.054}_{-0.051}$  \\
$\sqrt{e} \sin \omega$ & $0.001^{+0.097}_{-0.114}$  \\
$c_1 = 2 u_1 + u_2$ & $1.009^{+0.005}_{-0.005}$  \\
$c_2 = u_1 - 2 u_2$ & $-0.182^{+0.016}_{-0.014}$  \\
Occultation depth & $0.000031^{+0.000007}_{-0.000007}$  \\
 &    \\
\textit{Deduced stellar parameters} &    \\
\tableline
 &    \\ 
$u_1$ & $0.367^{+0.003}_{-0.003}$  \\
$u_2$ & $0.274^{+0.006}_{-0.006}$  \\
Density $\rho_{\star}$ [$\rho_{\odot}$] & $0.354^{+0.017}_{-0.008}$  \\
Surface gravity $\log g_{\star}$ [cgs] & $4.175^{+0.015}_{-0.011}$  \\
Mass $M_{\star}$ [$M_{\odot}$]& $1.166^{+0.051}_{-0.054}$  \\
Radius $R_{\star}$ [$R_{\odot}$]& $1.483^{+0.025}_{-0.029}$  \\
Age [Gyr] &  $4.0^{+0.3}_{-0.4}$ \\
 &    \\
\textit{Observed stellar parameters}  &  \\
\tableline
 &  \\
\te & $5947 \pm 100$ \\
$\mathrm{[Fe/H]}$ & $0.07 \pm 0.04$ \\
$V\rm{sin} $$i$ & $0.8 \pm 0.5$ km s$^{-1}$ \\ \\
\textit{Deduced planet parameters}  &  \\
\tableline
 &  \\
RV $K$  [m\,s$^{-1}$]  & $48.2^{+4.4}_{-4.3}$ \\
$b_{transit}$ [$R_{\star}$]  & $0.174^{+0.011}_{-0.011}$ \\
$b_{occultation}$ [$R_{\star}$]  & $0.174^{+0.011}_{-0.011}$ \\
$T_{occultation}$ - 2450000 [HJD]  & $5010.666^{+0.004}_{-0.003}$ \\
Orbital semi-major axis $a$ [AU]  & $0.0556^{+0.0007}_{-0.0007}$ \\
Orbital inclination $i$ [deg]  & $88.76^{+0.08}_{-0.08}$ \\
Orbital eccentricity $e$  &  $<0.01$ (1$\sigma$), $<0.09$ (3$\sigma$) \\
Argument of periastron $\omega$ [deg]  & $182^{+97}_{-98}$ \\
Density $\rho_{P}$ [g\,cm$^{-3}$]  &$0.111^{+0.011}_{-0.009}$ \\
Surface gravity log $g_{P}$ [cgs]  & $2.57^{+0.04}_{-0.04}$ \\
Mass $M_{P}$ [$M_{Jup}$]  & $0.431^{+0.041}_{-0.040}$ \\
Radius $R_{P}$ [$R_{Jup}$]  & $1.695^{+0.028}_{-0.032}$ \\

\enddata
\label{tab:params}
\end{deluxetable}

\begin{center}
\begin{deluxetable}{lccccccc}
\tabletypesize{\scriptsize}
\tablecaption{{\it Warm-Spitzer} observations. }
\tablewidth{0pt}
\tablehead{\colhead{Visit} & \colhead{AOR} & \colhead{Wavelength} & \colhead{Obs. Date (UT)} & \colhead{Select. points} & \colhead{Depth (\%)} & \colhead{Weighted. Avg. depth}& \colhead{$T_{\rm bright}$}}
\startdata
1 & 40251392 & 3.6  &  2010-09-06  & 1233 & $0.141^{+0.026}_{-0.021}$ &              -    & -             \\
3 & 40250880 & 3.6  &  2010-12-26  & 1151 & $0.130^{+0.026}_{-0.032}$ & $0.137\pm0.020$\% & $1597\pm160$ K \\
2 & 40251136 & 4.5 &  2010-09-15   & 1160 & $0.108^{+0.046}_{-0.034}$ &              -    & -             \\
4 & 40250624 & 4.5 &  2011-01-08   & 1212 & $0.129^{+0.039}_{-0.061}$ & $0.116\pm0.031$\% & $1429\pm190$ K \\
\enddata
\label{tab:spitzer}
\end{deluxetable}
\end{center}

\end{document}